\documentclass[9pt,twoside,twocolumn,letter]{article}

\usepackage[left=0.5in,right=0.5in,top=0.5in,bottom=1in]{geometry}  
\usepackage{graphics}
\usepackage{graphicx}

\usepackage{mathtools,stmaryrd,soul}
\usepackage{amsmath,amssymb}
\usepackage{authblk}

\usepackage[utf8]{inputenc}
\usepackage[title]{appendix}

\usepackage{caption}

\begin{document}

\title{Topological transformations of speckles}

\author[1]{Jérôme Gateau}
\author[2]{Ferdinand Claude}
\author[2,3]{Gilles Tessier}
\author[2,4,*]{Marc Guillon}

\affil[1]{Sorbonne Université, CNRS, INSERM, Laboratoire d’Imagerie Biomédicale, LIB, 15 rue de l'école de Médecine, F-75006, Paris, France}
\affil[2]{Neurophotonics Laboratory, CNRS UMR~8250, Paris Descartes University, Sorbonne Paris Cit\'e, Paris, France}
\affil[3]{Sorbonne Université, Institut de la Vision, INSERM UMR S968, CNRS UMR 7210, 17 rue Moreau, 75012, Paris, France}
\affil[4]{Saint-Pères Paris Institute for Neurosciences (SPPIN), CNRS UMR~8003, Paris Descartes University, Sorbonne Paris Cit\'e, Paris, France}

\affil[*]{Corresponding author: marc.guillon@parisdescartes.fr}

\maketitle

\bigskip 

\textbf{
Deterministic control of coherent random light is highly important for information transmission through complex media.
However, only a few simple speckle transformations can be achieved through diffusers without prior characterization.
As recently shown, spiral wavefront modulation of the impinging beam allows permuting intensity maxima and intrinsic $\pm 1$-charged optical vortices.
Here, we study this cyclic-group algebra when combining spiral phase transforms of charge $n$, with $D_3$- and $D_4$-point-group symmetry star-like amplitude modulations.
This combination allows statistical strengthening of permutations and controlling the period to be $3$ and $4$, respectively. Phase saddle-points are shown to complete the cycle.
These results offer new tools to manipulate critical points in speckles.}

\section{Introduction}

The propagation of coherent light through scattering media yields random wavefields with typical intensity structures called optical speckles. The control of light distribution inside and through complex media by wavefront modulation of the impinging beam is of critical importance for application ranging from bio-imaging~\cite{Yang_NP_15} to telecommunications~\cite{Padgett_OL_12}, for instance by multiplexing information with orbital angular momentum~\cite{Li_LSA_19}. Information transmission through diffusers is typically characterized in terms of field and intensity correlations~\cite{Akkermans_Montambaux}. For diffusers exhibiting so-called `'memory effect''correlations, important invariants were identified under specific spatial (tilt and shift) transformations ~\cite{Stone_PRL_88,Feng_PRL_88,Judkewitz_NP_15,Vellekoop_optica_17}. Additionally, regardless of the wavefront of the impinging beam, critical points in random wavefields exhibit many topological correlations~\cite{Freund_1001_correlations}, which thus demand the development specific tools to be analyzed. Optical vortices are especially important critical points since they are centered on singular phase points coinciding with nodal points of the intensity. They spontaneously appear in random wavefields~\cite{Berry_PRSLA_74}, and thereby allow efficient super-resolution microscopy~\cite{Pascucci_PRL_16,Pascucci_ArXiv_17}. The present work aims at exploring the possibility to manipulate topological correlations between critical points in random wavefields under symmetry control and spiral wavefront modulation, in a Fourier plane of the impinging beam.

Critical points are characterized by their topological charge and their Poincaré number~\cite{Dennis_thesis}. They may typically be controlled by applying phase or amplitude masks in a Fourier plane. Any smooth and regular transform of the wavefield (either in phase or amplitude) induces changes  preserving both the topological charge and the Poincaré number~\cite{Freund_1001_correlations,Nye_PRSLA_88}. Noteworthy, these conservation rules account for the topological stability of isolated vortices of charge $1$ in speckles since the creation or annihilation of vortices can only occur by pairs~\cite{Nye_PRSLA_88, Pascucci_PRL_16}. 
As opposed to smooth phase transforms, the addition of a spiral phase mask in a Fourier plane is a \emph{singular transform} and results in a change of the total orbital angular momentum~\cite{Larkin_JOSA_01,Padgett_PO_09}.
Recently, considering correlations between the spatial distribution of critical points in a speckle under such spiral phase transforms~\cite{Gateau_PRL_17}, we observed a strong inter-play between intensity maxima and optical vortices. More precisely, the obtained results suggested that the topological charge of these critical points were all incremented by applying a $+1$ spiral phase mask in the Fourier plane. The impossibility to spontaneously get $+2$-charged vortices (unstable and thus unlikely in random light structures~\cite{Freund_OC_93}) resulted in the observation of a partial cyclic permutation of the three populations of critical points (namely, maxima and $\pm 1$-charged vortices).
Furthermore, as a third kind of possible transform, it was observed that the orbital angular momentum may be not conserved when using amplitude masks with a high degree of symmetries~\cite{Cordoba_PRL_05,Xie_OL_12}. As a result, optical vortices can be created using simple amplitude masks~\cite{Visser_PRA_09,Cheng_AO_14}. This property proved to be of interest for imaging applications to reveal symmetries of an imaged object~\cite{Boyd_LSA_17,Willner_OL_17,Chen_OL_16} and for allowing topological charge measurements~\cite{Chavez-Cerda_PRL_10}, especially in astronomy~\cite{Beijersbergen_PRL_08}.  

Here, combining spiral phase transforms of order $n$ with star-like amplitude masks having discrete point group symmetries $D_3$ and $D_4$, we study experimentally the topological correlations between intensity maxima and optical vortices in speckles. A new co-localization criterion is proposed, inspired by statistical mechanics. Although random wavefields do not possess any symmetry, such a combination allows us to strengthen periodicity and even to control the period of the cyclic permutation. Noteworthy, for an amplitude mask of symmetry $D_4$, a phase saddle point appears as a complementary critical point to complete a cycle of period 4. A transposition between vortices of charge -1 and vortices of charge +1 is also revealed when adding a 2-charged spiral phase mask.

\section{Experimental procedure} 

The experimental procedure consisted in modulating a random phase pattern in a Fourier plane with an amplitude mask and a spiral phase mask. Here, spiral phase masks of order $n$, ${\rm SP}_{n}(\theta) =  e^{i.n.\theta}$ (in polar coordinates), were applied for $n \in \llbracket -6;6 \rrbracket$. As amplitude masks, three Binary Amplitude (${\rm BA}$) masks were used: a disk and two periodic angular slits with a point group symmetry $D_3$ and $D_4$. They are defined by the following angular transmission function (in polar coordinates): 
\begin{equation}
{\rm BA}^{\rm N}(\theta) = \begin{cases}
1, & \text{ if $\lvert \theta-(k-\frac{1}{2}).\frac{2.\pi}{\rm N} \rvert\ < \frac{\pi}{32}$ with $k \in \llbracket 1;\rm N \rrbracket$ }.\\  
0, & \text{otherwise}.
  \end{cases}
\label{eq:apert}
\end{equation} 
for $N\in\left\{3,4\right\}$. By convention, ${\rm BA}^{\infty}$ defines the disk-shaped aperture (obtained for $\rm N \geqslant 32$). For $\rm N < 32$, the orbital angular momentum content (or spiral spectrum~\cite{Torner2005}) of the ${\rm BA}^{\rm N}$ aperture exhibits discrete harmonics for the spiral modes $n= \rm{p.N}$ with ${\rm p}\in \mathbb{Z}$.
Given the width of the angular slits and provided that $N$= 3 or 4, the aperture ${\rm BA}^{N}$  can be considered as invariant by the addition of ${\rm SP}_{n}$ when $n = \pm \rm N$ or $\pm 2.\rm N$. Such an invariance in the Fourier plane is thus necessarily associated with a periodic transform of the speckle pattern in the real space. 

\begin{figure}[htb]
\centering
\fbox{\includegraphics[width=\linewidth]{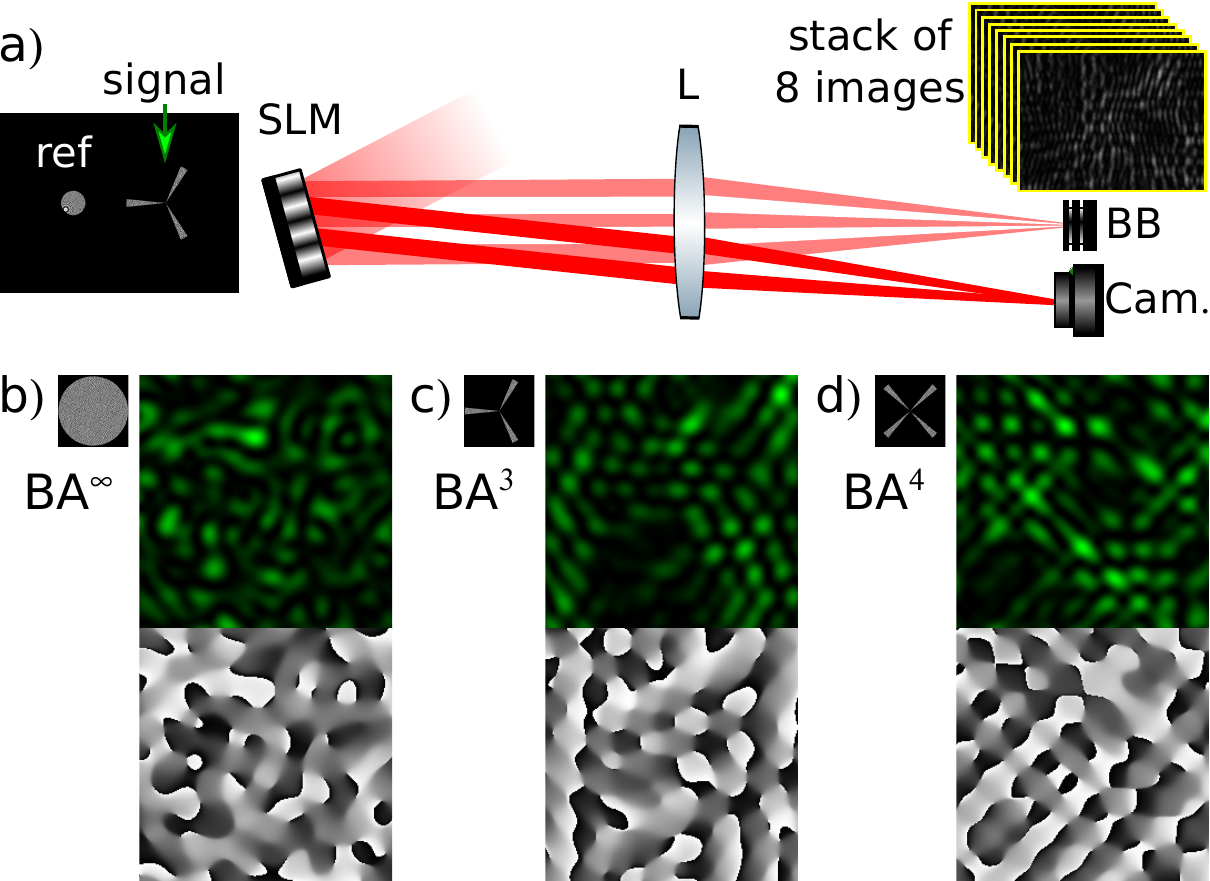}}
\caption{Experimental setup (a) used to measure the intensity and the phase of speckle patterns corresponding to the different binary amplitude masks. A spatial light modulator (SLM) is illuminated with a collimated laser beam at 635nm. The phase is measured by phase stepping interferometry. Both the reference and the signal wavefronts are imprinted on the SLM. in addition to a blazed grating which allows sending undiffracted light to a beam block (BB) and the first order diffracted beam to a camera (Cam.). A stack of eight images was then sequentially recorded while phase shifting the reference beam. The measured intensity $I_{0}$ maps (top, green colorscale) and phase $\Phi_{0}$ maps (bottom, gray colorscale) are presented for  a circular aperture (${\rm BA}^{\infty}$) (b), periodic angular-slits with a point group symmetry $D_3$ (${\rm BA}^{3}$) (c), and  periodic angular-slits with a point group symmetry $D_4$ (${\rm BA}^{4}$) (d). Miniatures of the ${\rm BA}$ masks are displayed for illustration.}
\label{fig:principle}
\end{figure}

The experimental configuration is detailed in Fig.~\ref{fig:principle} [See Supplement 1, Section 1 for further details on the experimental methods]. A spatial light modulator (SLM) (LCOS, X$10468$, Hamamatsu) was illuminated with a collimated laser beam at $635~{\rm nm}$ and Fourier conjugated to a camera (768x1024 pixels, pixel size: $\rm{4.65x4.65}$ $\mu \rm{m^2}$) with a converging lens. The phase $\Phi_{n}$ and amplitude $A_{n}$ of the modulated (${\rm SP}_{n}$ mask) random wave were measured at the camera plane by phase stepping interferometry~\cite{Creath_2005}. To do so, the SLM (792x600 SLM pixels, pixel size: $\rm{20x20}$ $\mu \rm{m^2}$) was split in two parts to generate both the modulated random wave (or signal wave) on one side and a reference wave on the other side (see Fig.~\ref{fig:principle}a). The signal wave was generated by adding simultaneously the scattering random phase pattern, the spiral phase modulation ${\rm SP}_{n}$ and the amplitude mask ${\rm BA}^{N}$. Adding a blazed grating achieved spatial separation of the imprinted signal wavefront from undiffracted light (the latter being sent to a beam-block). The signal speckle intensity $I_{n}$ could be measured directly by removing the contribution of the reference beam. 

For phase-stepping interferometry, an additional Fresnel lens was added to the reference beam in order to cover the camera surface. The latter spherical contribution as well as the relative phase-tilt between the signal and the reference beams were removed in a numerical post-processing step. A stack of eight images was sequentially recorded by phase shifting the reference beam by $2 \pi/8$ phase-steps. All BA masks had the same radius of $r=170$ pixels at the SLM, so yielding the same speckle grains size on the camera plane: $\lambda/(2{\rm NA})$ = 70 $\mu \rm m$ -Full Width Half Maximum (FWHM), where $\lambda$ is the wavelength and $\rm{NA}$ $\simeq r/f \simeq 4.53\times 10^{-3}$ the numerical aperture of illumination (with $f=750~{\rm mm}$ the focal length of the lens L in Fig.~\ref{fig:principle}a). The speckle grain size thus covered $15$ camera pixels and ensured a fine sampling of the speckle patterns. Hereafter, all distances and spatial densities are expressed setting $\lambda/(2{\rm NA})$ as the length unit.

In Fig.~\ref{fig:principle}(b-d), an illustration of the speckle intensity and phase maps obtained for the three different geometries of BA masks is shown. 
In the following study, intensity maps $I_{n}$ and phase maps $\Phi_{n}$ were measured for all three ${\rm BA}^{N}$ masks ($N\in \{ 3,4,\infty \}$) and for each ${\rm SP}_{n}$ masks ($n \in \llbracket -6;6 \rrbracket$). For comparison, the intensity-map $I_{\rm rand}$ and the phase-map $\Phi_{\rm rand}$ obtained for a non-correlated scattering random pattern were acquired for all the ${\rm BA}^{N}$ masks independently.

\section{Statistical analysis of the topological correlation between critical points}
\subsection{Studied critical points}

The field at the camera being linearly polarized, optical fields are here studied as scalar fields. The location of the main critical points of the experimental intensity and phase maps were measured [See Supplement 1, Section 1.C for details on the detection of critical points] and their statistical correlation distances were analyzed. Importantly, for a given BA mask, adding ${\rm SP}_n$ masks preserves all statistical properties of the speckle patterns, such as the number-density of critical points. Phase saddle-points of $\Phi_{n}$ are notated $S_{n}^{p}$ and vortices of charge $\pm 1$: $V_{n}^{\pm}$. Vortices of charge higher than $1$ do not appear in Gaussian random wavefields~\cite{Freund_OC_93}. Maxima and saddle-points of $I_{n}$ are notated $M_{n}$ and $S_{n}^{I}$, respectively. Non-zero minima and phase extrema have not been considered here, since having significantly lower densities~\cite{Freund_PLA_95}. All the notations are summarized in Table~\ref{tab:critical-points}. 

The measured average number-densities of the critical points are presented in Table~\ref{tab:density-points}. The density of the critical points of type $X$ ($X =$ $V^{\pm}$, $M$, $S^{p}$ or $S^{I}$) is notated $\rho(X)$. As expected, $\rho(V^{-})$ and $\rho(V^{+})$ are equal~\cite{Freund_1001_correlations}, and $\rho(X)$ depends both on the type of critical point and the BA mask.

\begin{table}[htbp]
\centering
\caption{\bf Notations for the main critical points}
\begin{tabular}{cccc}
\hline
\bf Phase & Maxima & Saddle & Vortices (charge $\pm 1$)  \\
 & - & $S^{p}$ & $V^{-}$ and $V^{+}$ \\ 
\hline
\bf Intensity & Maxima & Saddle & Zeros  \\
 & $M$ & $S^{I}$ & $V^{-}$ and $V^{+}$ \\ 
\hline
\end{tabular}
  \label{tab:critical-points}
\end{table}

\begin{table}[htbp]
\centering
\caption{\bf Measured average number density of critical points (length unit: $\lambda/(2.{\rm NA})$). The average number of $V^{-}$ is 660.85 for the circular aperture (${\rm BA}^{\infty}$). }
\begin{tabular}{ccccc}
\hline
BA mask & $V^{-}$(or $V^{+}$) & $M$ & $S^{p}$ & $S^{I}$  \\
\hline
${\rm BA}^{\infty}$ & $0.19$ & $0.32$ & $0.36$ & $0.65$ \\
${\rm BA}^{3}$ & $0.20$ & $0.39$ &$0.36$ & $0.79$ \\
${\rm BA}^{4}$ & $0.19$ & $0.33$ & $0.40$ & $0.70$ \\
\hline
\end{tabular}
  \label{tab:density-points}
\end{table}

\subsection{Statistical tools for the analysis of topological correlations}

\begin{figure}[h]
\centering
\fbox{\includegraphics[width=\linewidth]{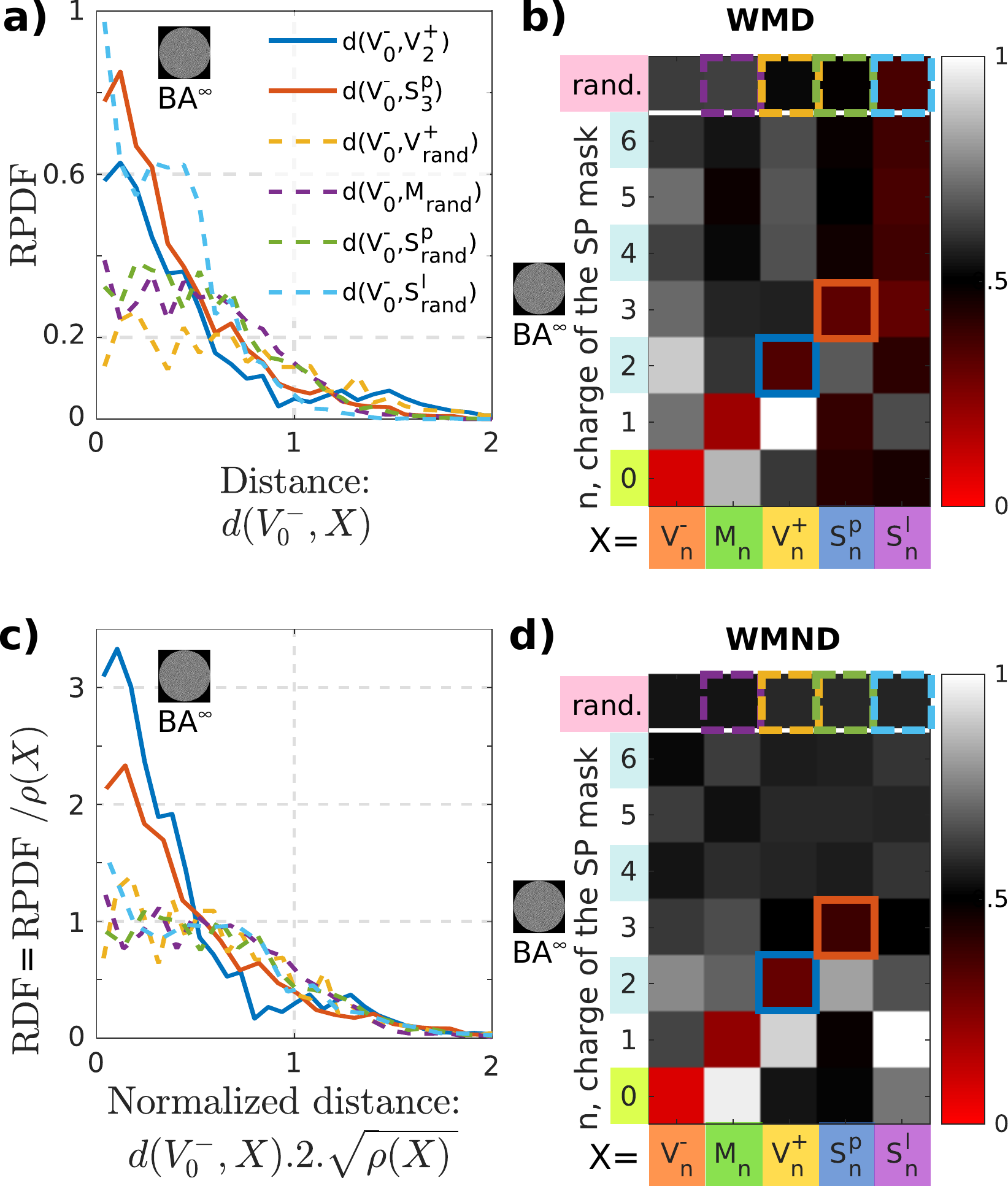}}
\caption{Statistical analysis of the separation distances between one set of critical points (here $V_0^-$) and the closest point of another set (notated $X$). The distance between $V_0^-$ and the closest $X$ is notated {\it d}$\left(V_0^-, X\right)$. The Radial Probability Density Functions (RPDF) of the nearest neighbor (a) and the corresponding Weighted Median Distance (WMD) (b) are shown. Radial Distribution Functions (RDF) of the nearest neighbor (c) and the corresponding Weighted Median Normalized Distance (WMND) (d) provide a statistical toolbox to study the spatial correlation between pairs of critical points. The results were derived from experimental measurements of $I_n$ and $\Phi_{n}$ obtained for the amplitude mask ${\rm BA}^{\infty}$.}
\label{fig:defnormalized}
\end{figure}

To study statistical transformations of critical points quantitatively, new specific tools are presented. What we discuss as transformation of critical points by the addition of spiral phase masks refers to the mean nearest neighbor distances between populations of critical points and calls for a discrimination parameter. Two specific statistical tools are then described below: the radial density function (RDF) and the Weighted Median Normalized Distance (WMND).

In our previous study~\cite{Gateau_PRL_17}, correlations between critical points could be characterized by computing the radial probability density function (RPDF) of the nearest-neighbor distance.
Fig.~\ref{fig:defnormalized}a presents RPDFs of the distance {\it d}$\left(V_0^-, X\right)$ in the case of ${\rm BA}^{\infty}$. We define {\it d}$\left(Y,X\right)$ as the distance between a $Y$-point and the closest $X$-point. The RPDF(r) corresponds to the probability to find the closest $X$-point at the distance $r$ from a $Y$-point, per unit area ($\int_0^\infty RPDF(r)2\pi r dr=1$). One drawback associated with the use of the RPDF is that it may suggest paradoxes if improperly interpreted. Considering {\it d}$\left(V_0^-, V_2^+\right)$ and {\it d}$\left(V_0^-, S_{\rm rand}^I\right)$, it seems that $V_0^-$ correlates both with $V_2^+$ and $S_{\rm rand}^I$ since both RPDFs reach high values at zero distances.
 
While a correlation is expected in the former case (due to topological charge incrementation), no correlation is expected from the latter which involves two independent sets of random points. The reason why the amplitude of the RPDF of {\it d}$\left(V_0^-, S_{\rm rand}^I\right)$ is higher than the one of {\it d}$\left(V_0^-, V_2^+\right)$ at zero distances, is just due to the $\sim 3$-times higher spatial density of intensity saddle points $S_{\rm rand}^I$ as compared to vortices $V_2^+$ (see Table~\ref{tab:density-points}): the probability to find a saddle point at close distance is thus larger. 
To quantitatively characterize topological correlations, we thus need to normalize RPDFs by the number densities $\rho(X)$. 

Our first statistical tool, the radial-distribution function (RDF) -- well known in statistical mechanics~\cite{chandler1987introduction} -- was extended here for nearest neighbor by normalizing the RPDF of {\it d}$\left(V_0^-, X\right)$ by $\rho(X)$, and the distances {\it d}$\left(V_0^-, X\right)$ by the mean X-interpoint half-distance $\left(2\sqrt{\rho(X)}\right)^{-1}$ ~\cite{Tournus2011}. Fig.~\ref{fig:defnormalized}c shows the RDF of the same data as in Fig.~\ref{fig:defnormalized}a. As a result, all the RDFs of {\it d}$\left(V_0^-, X_{\rm rand}\right).2.\sqrt{\rho(X)}$ are superimposed for every $X_{\rm rand}$, and the spatial correlation betweeen $V_0^{-}$ and $V_2^+$ clearly appears.

To obtain a single binary parameter discriminating the spatial correlation between $V_0^-$ and $X$, we further define the Weighted Median Normalized Distance (WMND) as a second statistical tool: the WMND$\left(V_0^-, X\right)$ is the $50\%$ weighted percentile of {\it d}$\left(V_0^-, X\right).2.\sqrt{\rho(X)}$ with weights corresponding to the RDF values. A ${\rm WMND}\left(V_0^-, X\right)$ around $0.5$ means that no spatial correlation exist between $V_0^-$ and $X$, while ${\rm WMND}<0.5$ and ${\rm WMND}>0.5$ mean an attraction and a repulsion, respectively. A zero WMND value means perfect correlation while WMND=1 means perfect anti-correlation. 

Fig~\ref{fig:defnormalized}d presents the WMND$\left(V_0^-, X\right)$ for all the critical points considered in this study and for $n\in \llbracket 0;6 \rrbracket$. For comparison, the Weighted Median Distance (WMD) associated with the RPDFs -- defined as the $50\%$ weighted percentile of {\it d}$\left(V_0^-, X\right)$ with weights corresponding to the RDPF values -- is also computed and displayed in Fig.~\ref{fig:defnormalized}b. 
To validate this tool, taking $BA^\infty$ as an illustrative example, we notice that WMND$\left(V_0^-, X_{\rm rand}\right)$ is around 0.5 for all the $X_{\rm rand}$, as expected. By comparison, WMD$\left(V_0^-, S^I_{\rm rand}\right)=0.35$, which irrelevantly suggests correlations as discussed above. Moreover, for $n>3$, the RDFs of {\it d}$\left(V_0^-, X_n\right)$ are observed to match the RDFs of {\it d}$\left(V_0^-, X_{\rm rand}\right)$: no noticeable spatial correlation is obtained for $n>3$. As expected again, the WMND$\left(V_0^-, X_{\rm rand}\right)$ are around 0.5 for $n>3$. Conversely, we get WMD$\left(V_0^-, S^I_{n}\right)< 0.38$, which would falsely suggest correlations. All these observations validate the WMND as a parameter to assess the spatial correlation between pairs of critical points in a speckle pattern.

\section{Topological correlations between critical points for the different amplitude masks}

\begin{figure}[htb]
\centering
\fbox{\includegraphics[width=\linewidth]{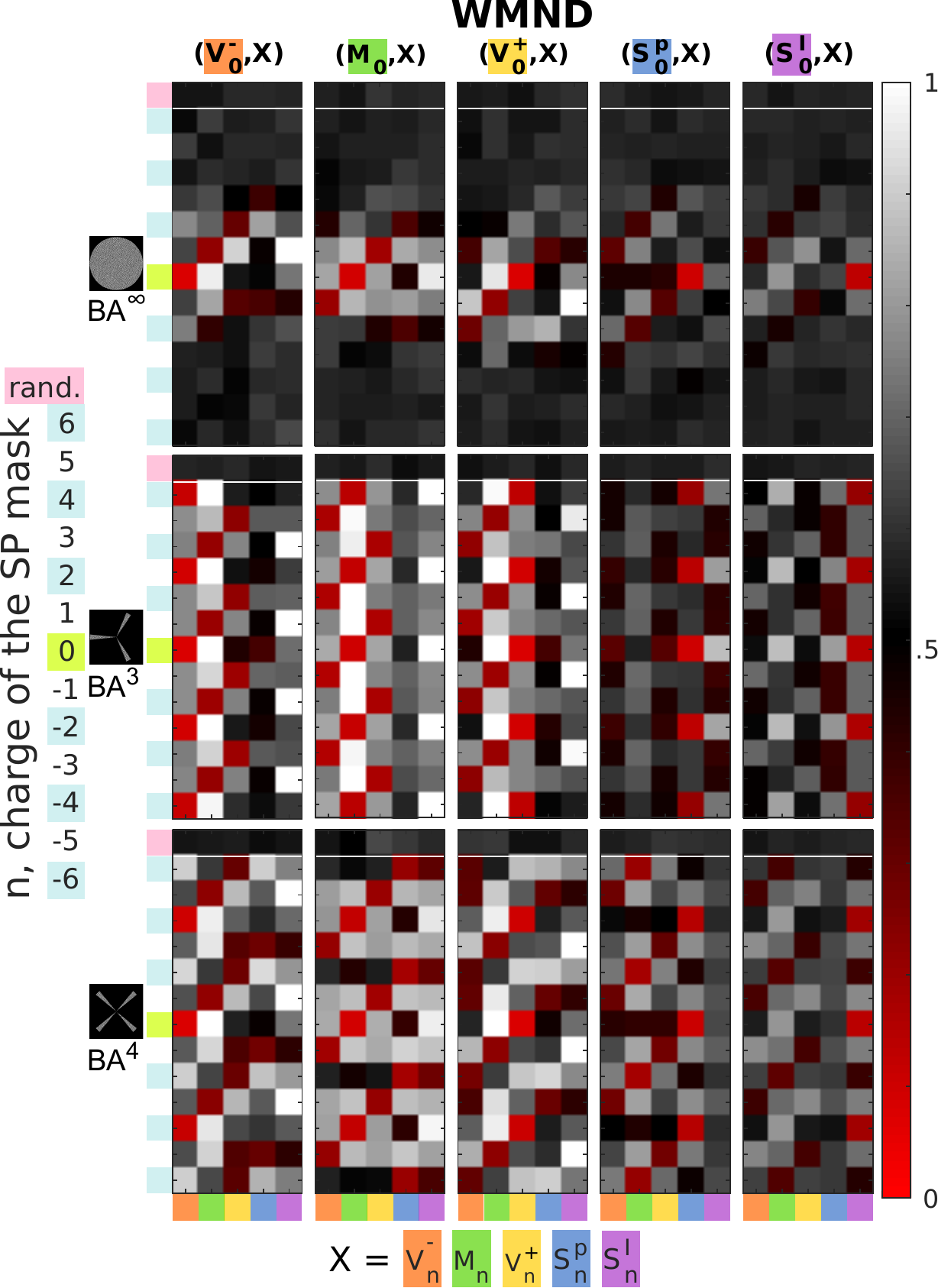}}
\caption{Weighted Median Normalized Distance (WMND) for all possible pairs of critical points screened here and for the addition of spiral phase mask with charges up to n =$\pm 6$. The WMND were computed from experimental measurements of $I_n$ and $\Phi_{n}$. }
\label{fig:multiplesp}
\end{figure}

Fig.~\ref{fig:multiplesp} presents the WMND($\rm{Y}_0$, $X_n$) for all the critical points ($\rm{Y}_0$ and $X_n$) screened in this study, for ${\rm SP}_{n}$ masks with $n \in \llbracket -6;6 \rrbracket$ and for the three considered BA apertures. The WMND was verified to be around $0.5$ for all the amplitude masks and all the pairs ($\rm{Y}_0$,$X_{\rm{rand}}$), for which there is obviously no spatial correlation. For the sake of readability, in the following, we only discuss the interplay between critical points when adding positively charged $\rm SP$ masks but symmetrical behaviours are observed for negatively charged $\rm SP$ masks (Fig.~\ref{fig:multiplesp}).

For the aperture ${\rm BA}^{\infty}$, the WMND reveals several noticeable features (reported in Table~\ref{tab:transformation}). First, as expected from our previous study~\cite{Gateau_PRL_17}, we notice some spatial correlations for the triplets $(V_{m-1}^-, M_m , V_{m+1}^+)$.  
Because $\rho(V_0^-)<\rho(M_1)$, a one-to-one transformation is impossible between vortices and maxima. 
Although $V_0^-$ and $V_2^+$ have the same number density, we also notice that WMND$(V_0^-,V_2^+)$ is significantly different from $0$, indicating that the rate of the macroscopic transformation from $V_0^-$ to $V_2^+$ is below 1.

In agreement with the 3-point cyclic permutation algebra observed in~\cite{Gateau_PRL_17}, a weak attraction is found for the pair $(V_0^{+}, V_{+1}^{-})$, corresponding to the topological-charge equation $1+1=-1$. However, as an alternative transformation for $V_0^+$, a similar attraction is now also observed for $(V_0^{+}, S^p_{+1})$.
$V_0^+$ is thus subjet to a bifurcation between $V_{+1}^-$ and $S^p_{+1}$, which implies two different mechanisms.

As a first possibility, some $V_0^{+}$ transform into $S^p_{+1}$. This transformation inspires the following interpretation: When adding a ${\rm SP}_{+1}$ phase mask to an isolated Laguerre-Gaussian beam with topological charge $+1$, a $+2$-charged vortex is obtained. Under weak perturbation, this $+2$ vortex splits into two $+1$ vortices, accompanied by the creation of both an intensity saddle point and a phase saddle point in between~\cite{Nye_PRSLA_88}. The creation of this pair of saddle points is governed by the Poincaré number conservation. In the frame of this model, a $V_0^+$ vortex is expected to co-localize with both $S_{+1}^I$ and $S_{+1}^p$ and to anti-correlate with the two $V_1^+$ that split away. Although no noticable spatial attraction was found for $(V_0^+,S_{+1}^I)$, a co-localization is observed for $(V_0^{+},S^p_{+1})$ and a weak repulsion is observed for the pair $(V_0^{+}, V_{+1}^{+})$, consistent with this interpretation. In speckles, where $+2$-charged vortices cannot be encountered since unstable~\cite{Freund_OC_93}, the weak perturbation approximation cannot be fully valid, potentially accounting for the remaining discrepancy between experimental observations and the proposed model.

As a second possible transformation, the more surprising attraction of the pairs $(V_0^{+}, V_{+1}^{-})$ is observed, which calls for another mechanism. Since no such transformation can be imagined from isotropic $V_0^{+}$, it may only be interpreted by a mechanism dominated by strong perturbations. The statistically uniform mesh created by vortices and maxima in speckles~\cite{Longuet_Higgins_JOSA_60}, together with strong correlations observed for pairs $(M_0,V_{+1}^{+})$ and $(V_0^{-},M_1)$ seem to constrain $V_0^{+}$ to co-localize with $V_{+1}^{-}$. This transformation would deserve further analytical investigation but we anticipate that the creation mechanism of $V_{+1}^{-}$ from $V_0^{+}$ can only be a many-body problem, involving the field structure (maxima, phase saddles and vortices) surrounding the initial $V_0^{+}$ of interest.

When adding a $\rm{SP}_{+2}$ mask for ${\rm BA}^\infty$, $V_0^+$ is not observed to significantly co-localize with any remarkable critical point (see Fig.~\ref{fig:multiplesp} and Table~\ref{tab:transformation}), whereas two possible transformations might have been expected for $V_0^+$. On the one hand, from the 3-point cyclic permutation, we could expect that $V_0^{+}$ would transform into $M_{+2}$. On the other hand, since in Table~\ref{tab:transformation}, maxima and phase saddle-points are noted to be simply exchanged (see pairs $(M_0,S^p_{+ 2})$ and $(S^p_0,M_{+ 2})$), a similar symmetrical echange between $-1$ and $+1$ vortices could be expected, yielding a transformation of $V_0^{+}$ into $V_2^-$ (as $V_0^{-}$ is transformed into $V_2^+$). However, no such correlation is observed either for the pair ($V_0^{+}$,$M_2$) or for ($V_0^{+}$,$V_2^-$). Conversely, these correlations appear when applying amplitude masks ${\rm BA}^3$ and ${\rm BA}^4$, respectively, as detailed in the following.
 
For $|n|>3$, no significant spatial correlation with the addition of $\rm{SP}_{n}$ is observed for ${\rm BA}^{\infty}$. This aperture has a circular symmetry. Therefore, its spiral spectrum contains only the fondamental spiral mode $n=0$, and is not invariant by the addition of any SP masks. All the described topological correlations associated with ${\rm BA}^\infty$ are summarized in Table~\ref{tab:transformation}.

\begin{table}[htbp]
\centering
\caption{\bf Macroscopic transformations observed for the critical points $\rm Y_0$ with the amplitude mask ${\rm BA}^{\infty}$. The transformation rates are below 1. }
\begin{tabular}{ccccc}

\hline
\bf  \bf Critical point $\rm Y_0$ & & \bf Adding ${\rm SP_{+1}}$ & & \bf Adding ${\rm SP_{+2}}$  \\
\hline
$V^{-}_0$ & $\rightarrow$& $M_1$ & $\rightarrow$ & $V^{+}_2$ \\
\hline
$M_0$ & $\rightarrow$& $V^{+}_1$ & $\rightarrow$ & $S^p_2$ \\
\hline
$V^{+}_0$ & $\rightarrow$& $V^{-}_{1} + S^p_{1}$ & $\rightarrow$ & $\emptyset$  \\
\hline
$S^{p}_0$ & $\rightarrow$& $V^{-}_1$ & $\rightarrow$ & $M_2$ \\
\hline \hline
\bf Adding ${\rm SP_{-2}}$ & & \bf Adding ${\rm SP_{-1}}$ & & \bf Critical point $\rm Y_0$  \\
\hline
$\emptyset$ & $\leftarrow$& $V^{+}_{-1} + S^p_{-1}$ & $\leftarrow$ & $V^{-}_0$ \\
\hline
$S^p_{-2}$ & $\leftarrow$& $V^{-}_{-1}$ & $\leftarrow$ & $M_0$ \\
\hline
$V^-_{-2}$ & $\leftarrow$& $M_{-1}$ & $\leftarrow$ & $V^{+}_0$  \\
\hline
$M_{-2}$ & $\leftarrow$& $V^{+}_{-1}$ & $\leftarrow$ & $S^{p}_0$ \\
\hline

\end{tabular}
  \label{tab:transformation}
\end{table}

As a possible solution to strengthen the 3-point cyclic permutation, we used the ${\rm BA}^3$ amplitude mask, making the Fourier plane almost invariant with respect to the addition of $\rm SP_{\pm 3k}$ (so long as $3k$, with $k$ integer, remains small enough). Here, four main observations can be noted.
First, as expected, the pairs $\left(\rm{Y}_0, \rm{Y}_{\pm 3}\right)$ and $\left(\rm{Y}_0, \rm{Y}_{\pm 6}\right)$ are observed to have a WMND very close to zero, indicating that the macroscopic transformation rate is close to $1$ for all these pairs. Second, we observe that the cycle of period 3 reinforces the spatial correlations of the triplet $(V_{m-1}^-, M_m , V_{m+1}^+)$ and even extends it to the $3^{rd}$ and $6^{th}$ spiral harmonics. Third, no noticeable correlation is observed between $V_{0}^{+}$ and phase saddle-points $S^p_{+ 1}$ (although an anti-correlation is obtained between $V_{0}^{+}$ and $V_{+1}^{+}$), contrary to the case of ${\rm BA}^{\infty}$. The periodicity of $3$ induces a strong correlation for the pairs $(V_0^{+}, V_{+1}^{-})$, and establishes a cyclic permutation of three populations of critical points $V^-$, $M$ and $V^+$. Forth, as a consequence, the pairs $(V_0^{+},M_{2})$ also exhibit strong spatial correlations, contrary to the case of ${\rm BA}^\infty$. In these two latter permutations, it must be reminded that not all intensity maxima $M$ may transform into vortices, because of the difference in spatial densities (Table~\ref{tab:density-points}) of these two populations of critical points~\cite{Gateau_PRL_17}.

Next, we constrained the period to be equal to $4$ by using the ${\rm BA}^{4}$ mask. In this case, the WMND$\left(\rm{Y}_0, \rm{Y}_{\pm 4}\right)$ are close to zero (transformation rate close to 1). As expected, this periodicity enhances the spatial correlation for the quadruplet  $( V_{m-1}^-, M_{m} , V_{m+1}^+ , S^p_{m+2})$ and extends it to their $4^{th}$ spiral harmonics. 
Furthermore, in Fig.~\ref{fig:multiplesp}, strong correlations are observed for the pairs $(V_0^-,M_1)$, $(M_0,V_{+1}^+)$ and $(S_0^p,V_{+1}^-)$. However, $V_0^{+}$ is still observed to bifurcate between $V_{+1}^{-}$ and $S_{+1}^{p}$ with the same likelihood, similarly to the ${\rm BA}^{\infty}$ case.
Therefore the cyclic permutation of four populations of critical points $S^p$, $V^-$, $M$ and $V^+$ is not clearly established for $\rm BA^4$ (conversely to the permutation obtained for $\rm BA^3$). 
 
When considering the addition of $\pm 2$-charged spiral masks for ${\rm BA}^{4}$, a transposition (2-cycle permuation) between $V^+$ and $V^-$ is clearly obtained (with a transposition rate below 1). A strong spatial correlation, reinforced as compared to ${\rm BA}^\infty$, is also clearly observed between $M$ and $S^P$ by addition of $\rm{SP}_{\pm 2k}$.

In summary, the results displayed in Fig.~\ref{fig:multiplesp} reveals the fondamental topological transformations of critical points in a speckle with the addition of $\rm SP$ mask for a single-spiral mode aperture (${\rm BA}^{\infty}$ ), and demonstrate the possible modification of topological transformation by the addition of BA masks with dihedral symmetry [See Supplement 1, Section 2.B for numerical confirmation of the experimental results and Section 2.C for details on transformations induced by $\rm BA$ masks with dihedral symmetries of orders higher than 4]. For the sake of simplicity, we chose here star-like amplitude masks with a dihedral symmetry which comprise spiral harmonics with equal amplitudes (at small enough spiral mode number) [See Supplement 1, Section 2.D for details on the influence of the width of the star branches]. 

\section{Wavefield control in the vicinity of critial points}

On a local scale, the transformations of critical points shown in Fig.~\ref{fig:multiplesp} arise from the convolution in the imaging plane of the scattered field with the point spread function (PSF) associated with the combined amplitude and spiral phase masks. Controlling the transformation of a critical point ${\rm Y}_0$ of the complex wavefield $A_0$, by adding a $\rm SP_n$ mask $(n \in \mathbb{Z}^* )$ in a Fourier plane, requires that the PSF associated with the combination of the $\rm SP_n$ and $\rm BA$ masks has a significant amplitude in the coherence area surrounding the critical point, or in other words: the area where the randomness of the speckle pattern has a limited influence as compared to the control by the incident wavefield. As a definition for the coherence area, we use the one proposed by Freund~\cite{Freund_94} : $C_{area}= (\rho(V^+)+\rho(V^-))^{-1}/2$. This definition avoids issues related to the shape of the aperture~\cite{Freund_94} encountered when considering the area of the intensity autocorrelation peak~\cite{Ochoa_83}. In our case, for all three $\rm BA$ masks, the coherence length was measured to be: $C_{length} = \sqrt{C_{area}} \simeq \lambda/(2.\rm{NA})$.

\begin{figure}[htb]
\centering
\fbox{\includegraphics[width=\linewidth]{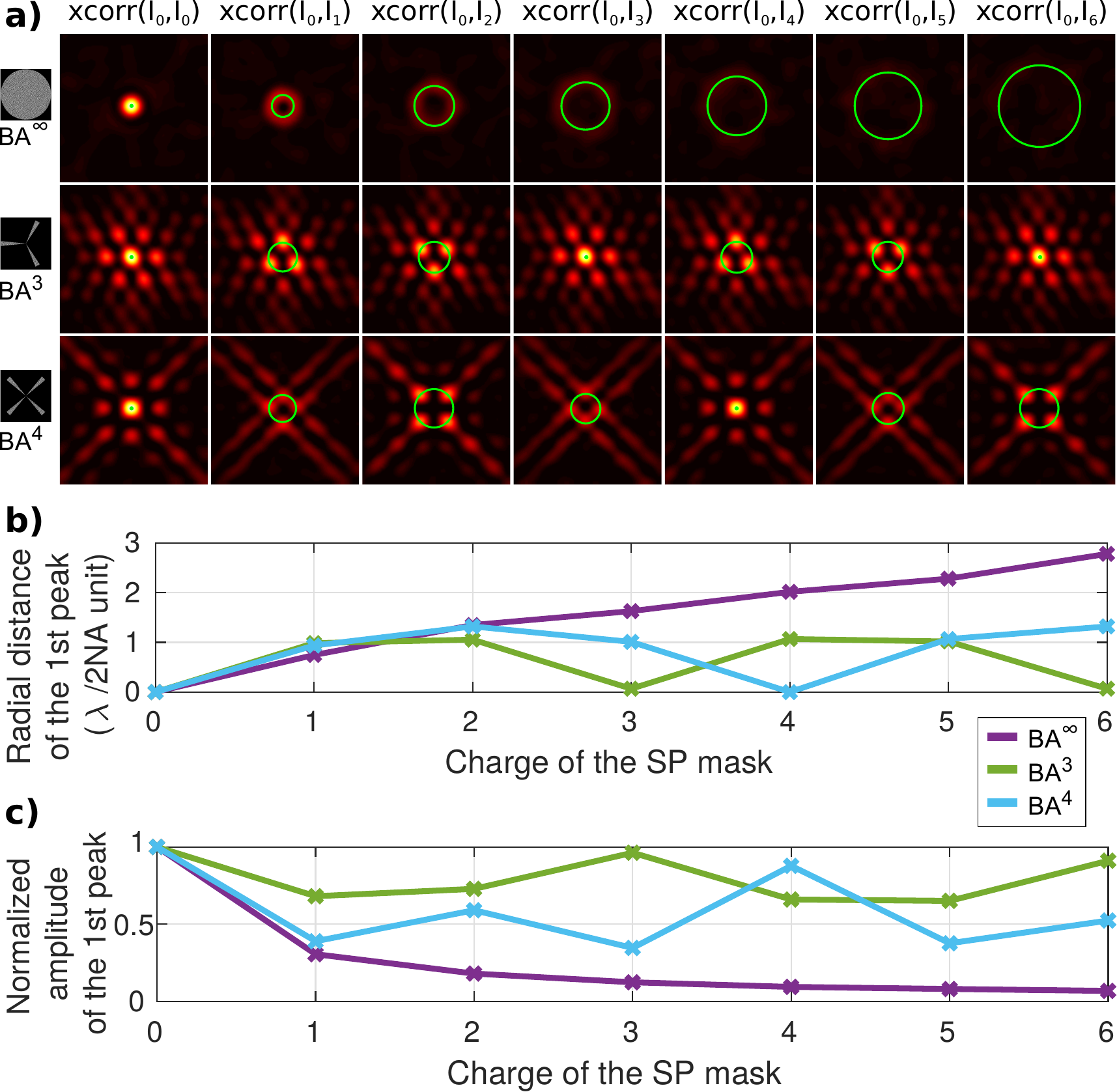}}
\caption{ Cross-correlation (a) of the speckle patterns $I_0$ and $I_n$ with $n\in \llbracket 0;6 \rrbracket$. The cross correlations illustrate the spatial distribution of the intensity in the point spread function associated with the combined amplitude and spiral phase masks. The colorscale set for the autocorrelations xcorr($I_0$,$I_0$) and was kept for the intercorrelations xcorr($I_0$,$I_n$). A N-periodicity is observed for the aperture $BA^{\rm N}$ with $\rm N$ = 3 or 4. Green circles mark the radial distance to the strongest peak. Radial distance (b) and normalized amplitude of the strongest peak (c) as a function of the charge of the spiral phase mask. The radial distance keeps increasing for the circular aperture ${\rm BA}^{\infty}$ while it remains close or below $\lambda/(2.\rm{NA})$ (the coherence length of the speckle) even for ${\rm{SP}_n}$ with $n\geqslant 3$ for the aperture ${\rm BA}^{\rm N}$ and $\rm N$ = 3 or 4. The amplitude decreases for ${\rm BA}^{\infty}$ while it remains above 0.3 for ${\rm BA}^{3}$ and ${\rm BA}^{4}$. }
\label{fig:intercorr}
\end{figure}

Experimentally, the PSFs can be obtained (Fig.~\ref{fig:intercorr}a) by computing the intensity cross-correlations of the measured speckle pattern $I_0$ and the measured speckle patterns $I_n$ associated with $\rm SP_n$ masks ($n \in \llbracket 0;6 \rrbracket$) and for the three $\rm BA$ masks. The mean values of $I_n$ were subtracted before computing the cross-correlations. The intensity cross-correlations xcorr($I_0$,$I_n$) are identical to the PSF of the combined $\rm BA$ and $\rm SP_n$ masks. The centered spot of the autocorrelation xcorr($I_0$,$I_0$) illustrates the spatial extent of the coherence area, and has the same dimension for all three $\rm BA$ masks since having the same radial aperture.

For ${\rm BA}^{\infty}$, we observe that xcorr($I_0$,$I_n$) has a circular symmetry with the highest values distributed on a ring whose radius (marked with a green line) increases with $n$ (Fig.~\ref{fig:intercorr}b), as observed for simple Laguerre-Gaussian beams~\cite{Grier2003}. Interestingly, for $n>3$, not only we observe that the ring radius is  larger than twice $C_{length}$ but also that its amplitude is decreased to below $1/10$ of the auto-correlation peak value (Fig.~\ref{fig:intercorr}c). As a consequence, the transformation of critical points by applying $\rm SP_n$ masks is inefficient (or ``unlikely'') and dominated by the surrounding random field. For this reason, no spatial correlation between pairs of critical points $({\rm Y}_0,X_n)$ could be found for $n>3$. For $\rm BA^3$ and $\rm BA^4$, the cross-correlation patterns xcorr($I_0$,$I_n$) have dihedral symmetries $D_3$ and $D_4$ and a periodicity of $\rm N$ = 3 and 4, respectively. In both cases, the radial distance of the strongest peak remains below $1.4* C_{length}$, and its amplitude always remains above $1/3$ of the auto-correlation maximum value (Fig.~\ref{fig:intercorr}c). The addition of the $\rm SP_n$ mask thus allows controlling the field inside the coherence area surrounding critical points ${\rm Y}_0$, even for $n>3$. 

\section{Conclusion}

The critical points that naturally appear in a random wavefield can be transformed by the addition of a spiral phase mask in a Fourier plane. Here, we studied these transformations experimentally by imprinting spiral phase masks with a charge $n\in \llbracket -6;6 \rrbracket$ to a laser beam impinging on a randomly scattering surface. In addition, these phase masks were combined with star-like amplitude masks with dihedral symmetries $D_3$ and $D_4$ in order to better control critical point transformations.

For a simple disk-shaped aperture carrying a single spiral mode $n=0$, we experimentally demonstrated the topological correlation existing between the critical points of the initial wavefield $A_0$, and the corresponding spiral transformed field $A_n$. A partial transformation of vortices $V_0^{-}$ into maxima $M_{+1}$ was observed as well as a transformation of maxima $M_{0}$ into vortices $V_{+1}^{+}$. Vortices $V_0^{+}$ were observed to either correlate with phase saddle points $S_{+1}^{p}$ or with vortices of opposite sign $V_{+1}^{-}$. For this statistical bifurcation, two transformation interpretations were suggested, calling for further future analytical studies. No simple topological correlation was found between the critical points of the wavefields $A_0$ and $A_n$ for $|n|>3$. This result could be explained by the weak influence of spiral phase masks with a charge higher than $2$ in the coherence area surrounding the critical points. 

Furthermore, adding centered binary amplitude masks with dihedral symmetry $D_3$ or $D_4$ and \emph{Dirac-comb}-like spiral spectra (of period $3$ and $4$), we demonstrated that it is possible to deeply modify the topological correlation between critical points. The observed changes arise from the introduction of a periodicity in the transformation between the critical points. We could thereby extend the correlation to spiral phase masks with charges higher than 3, and reinforce some spatial correlation intrinsically present with a circular-aperture symmetry. For the amplitude mask with a $D_3$-symmetry, a cyclic permutation between negatively charged vortices $V^-$, maxima $M$, and positively charged vortices $V^+$, is observed. For the amplitude mask with a $D_4$-symmetry, phase saddle points participate as complementary points to complete the $4$-periodic cycle. Considering the addition of 2-charged spiral masks, transpositions between $V^-$ and $V^+$, and between $M$ and $S^p$ were also revealed for $D_4$-symmetry. The enhancement of the spatial correlation between the critical points of the wavefields $A_0$ and $A_n$ (compared to ${\rm BA}^{\infty}$) could be explained by the strong influence of the spiral phase mask in the coherence area surrounding each critical point, when the binary amplitude masks are added. 

Here, cyclic permutations were controled using binary amplitude masks enforcing periodicity. Interestingly, our study may extend to other amplitude masks with a dihedral symmetry, such as polygonal~\cite{Xie_OL_12} and triangular apertures~\cite{Chavez-Cerda_PRL_10}, whose interactions with vortex beams were studied in free space. For N-gons, the $\rm N^{th}$ spiral harmonics have a much lower amplitude than the fundamental spiral mode $(n=0)$. As a result, the spatial correlations between critical points for $|n|>3$ is weaker than for ${\rm BA}^{\rm N}$, and vanishes with the increasing charge of the $\rm SP$ mask [See Supplement 1, Section 2.E].

In a nutshell, we showed here that it is possible to manipulate the topological correlation between critical points and to control the transformation of critical points in random wavefields by combining amplitude masks and spiral phase transforms. Topological manipulation of critical points in random wavefields is of high importance to understand and control light propagation through scattering and complex media. The statistical study of correlations between permuted critical points provides a new tool to analyse seemingly information-less and random intensity patterns and thus to transmit information through complex media.

\section*{Funding Information}
This work was partially funded by the french Agence Nationale pour la Recherche (NEOCASTIP ANR-CE09-0015-01, SpeckleSTED ANR-18-CE42-0008-01).

\clearpage
\appendix

\part*{Supplementary information}

\textbf{The first section provides expended descriptions of the methods used to acquire and process the experimental data: details are given on the phase mask displayed on the spatial light modulator, the numerical treatment of the recorded speckle parterns is presented, and the algorithms for the detection of critical points are given. The second section contains numerical confirmations of the experimental results, as well as an extention (in numerical simulation) of the range of tested binary amplitude apertures: the Weighted Median Normalized Distances are shown for star-like amplitude masks with dihedral symmetries of orders higher than 4, for star-like amplitude masks with larger angular slits, and for polygonal apertures.}

\section{Experimental data acquisition and processing}

\subsection{Phase mask on the SLM}

\begin{figure*}[htb]
\centering
\fbox{\includegraphics[width=\linewidth]{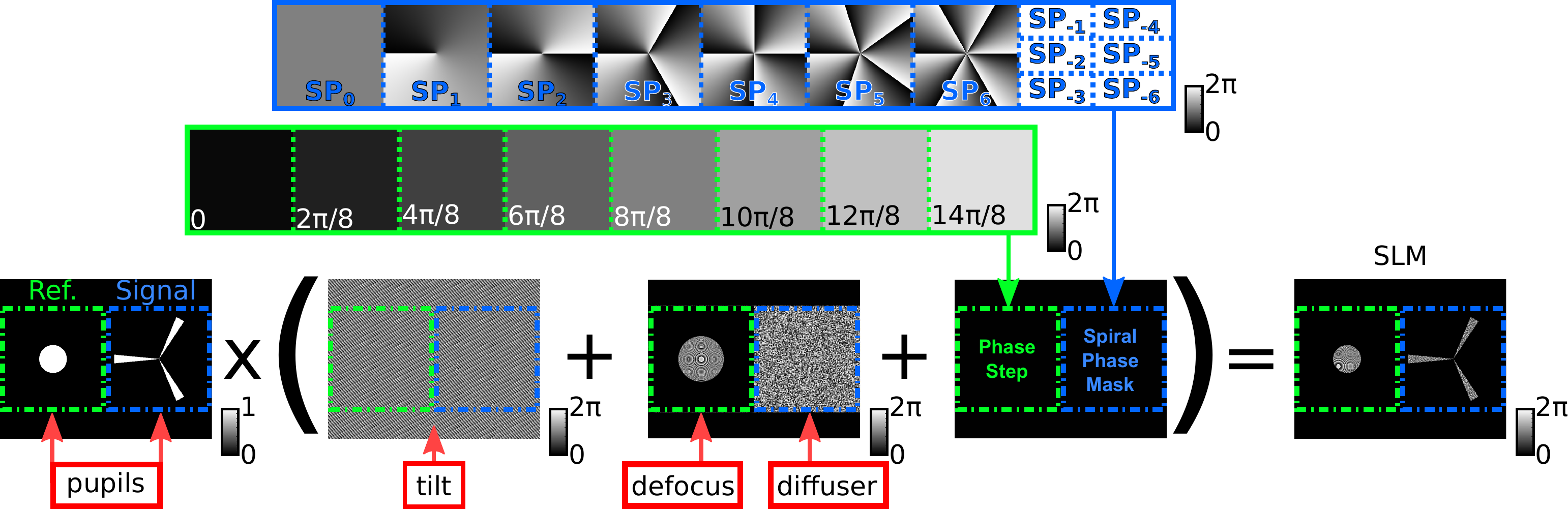}}
\caption{Computation of the phase displayed at the SLM. The SLM is virtually split into two parts of surfaces $396\times396$ pixels, each bearing pupils of different shapes, one for the reference beam (green), the other for the signal beam (blue). A tilt is applied to both beams in order to shift the speckle pattern of interest away from the unmodulated light. A defocus is added to the reference beam to cover the caerma surface, while the signal beams receives a random phase mask (or diffuser). Finally, the phase step and spiral wavefronts are added.}
\label{fig:phase_SLM}
\end{figure*}

The laser source was a laser diode emitting at $\lambda = 635~{\rm mm}$. The laser beam was spatially filtered with a pinhole, expanded and collimated to cover the surface of the Spatial Light Modulator (SLM). The phase at the SLM was computed according to the scheme shown in Fig.~\ref{fig:phase_SLM}.

The scattering phase pattern at the SLM (or diffuser in Fig.~\ref{fig:phase_SLM}) was engineered in order to both generate a fully developed speckle pattern at the camera plane and to minimize energy losses by controling the scattering angle. To achieve so, a complex matrix of dimension $396\times 396$ pixels (half of the long-axis of the SLM) was generated numerically, with uniform amplitude and random phases evenly distributed between $0$ and $2\pi$. This random matrix was then multiplied by a disk-shaped amplitude mask of radius $60$ SLM pixels (SLM pixel size : $px_{SLM} ^2 = 20\times20$~$\mu \rm m^2$) and Fourier transformed numerically to generate the random phase mask at the SLM plane. Provided the optical design of the setup and the focal distance of the lens $f = 750$~mm, this mask corresponds to a disk of diameter $\frac{60}{400}\times \frac{2.\lambda. f}{px_{SLM}}\simeq 8.8$~mm, slightly larger than the camera diagonal ($7.62$~mm), so ensuring uniform illumination of the camera chip (on average) and minimization of energy losses.

The $\rm BA$ amplitude mask were applied to the phase mask by setting the SLM phase to zero where the amplitude of the $\rm BA$ mask equals zero, and by adding a blazed grating where the amplitude of the $\rm BA$ mask equals one. The radius of the $\rm BA$ masks was chosen to have a speckle grain size of 15 camera pixels (camera pixel size: $4.65\times4.65$~$\mu \rm m^2$). The blazed grating had a period of 4.34 SLM pixels (SLM pixel size : $20\times20$~$\mu \rm m^2$), which yields a deflection angle of $7.31 ~\rm mrad$.  With the focal distance of the lens $f = 750$~mm, this deflection angle provides a 5 mm shift of the zero order as compared to the speckle pattern. Because of the camera field of view ($4.76~{\rm mm}\times 5.95~{\rm mm}$), the undiffracted zero-order was centered at a distance $2.5$~mm away from the camera sensor and could then efficiently be blocked with no spurious light observed on the camera. 

For the reference beam, the same blazed grating was applied to a disk-shaped aperture separated from the scattering aperture. The radius of the disk was chosen so that the signal beam (scattered wavefield) and the reference beam have similar amplitudes at the camera plane. A defocus (Fresnel lens) was applied to the reference beam to cover the camera surface. 

\subsection{Numerical treatment of experimental data}
The complex wavefield $A_n e^{i\Phi_n}$ at the camera plane was measured with the following procedure. The phase $\Phi_n$ was obtained thanks to the intensity modulation induced by the phase-stepping interferometric measurements~\cite{ROBINSON198626,CREATH2005364}. The phase-stepping was performed by acquiring eight successive images with relative phase-shifts between the signal and reference waves: $k\times2\pi/N$, with $N=8$ and $k \in \llbracket 0;7 \rrbracket $. The phase of the speckle pattern was obtained from phase-stepped intensity measurements: 
\begin{eqnarray}
I_{n,k}&=&\left|E_Re^{\frac{2ik\pi}{N}}+A_n e^{i\Phi_n}\right|^2\\
&=&\left|E_R\right|^2+ \left|A_n\right|^2+ E_R A_n e^{-i\Phi_n} e^{\frac{2ik\pi}{N}}+ E_R^\ast A_n e^{i\Phi_n} e^{-\frac{2ik\pi}{N}}\nonumber
\end{eqnarray} 
The phase can then be trivially retrieved by computing the argument of the following sum:
\begin{equation}
\displaystyle{\sum_{k=0}^{N-1}}I_{n,k}\times e^{\frac{2ik\pi}{N}}=E_R^\ast A_n e^{i\Phi_n}
\end{equation}
In our case, the reference beam did not exhibit a flat phase but a combination of a parabolic curvature as well as a relative phase tilt with respect to the speckled beam. Both of these profiles were numerically removed from the computed phase $\Phi_n$. The magnitude $A_n$ was not computed using this result to avoid amplitude uncertainties about the reference beam. Instead, $A_n$ was obtained from the signal speckle intensity $I_n$ without any reference beam: $A_n = \sqrt{I_n}$.

For all the apertures and all the speckle patterns, the Full Width Half Maximum (FWHM) of the point spread function (i.e. the speckle grain size) corresponded to 15 camera pixels, leading a fine spatial sampling of the speckle field. We set the length unit to 15 pixels and the corresponding spatial sampling frequency to 15 $\rm pixel^{-1}$. For noise removal, the complex amplitude field was filtered in the Fourier domain by zeroing the values corresponding to spatial frequencies outside of the disk $\sqrt{ f_x^2 + f_y^2} \leqslant 0.5$ $\rm pixel^{-1}$. Because of the spatial filtering by optical system, the frequencies outside of this disk contain only optical and electronic noise. We verified that the strongest spectral values of the complex field were within the disk. After this filtering process, experimental noise was not found to interfere with the detection of the critical points. 

\subsection{Detection of the critical points} 

\begin{figure}[htb]
\centering
\fbox{\includegraphics[width=\linewidth]{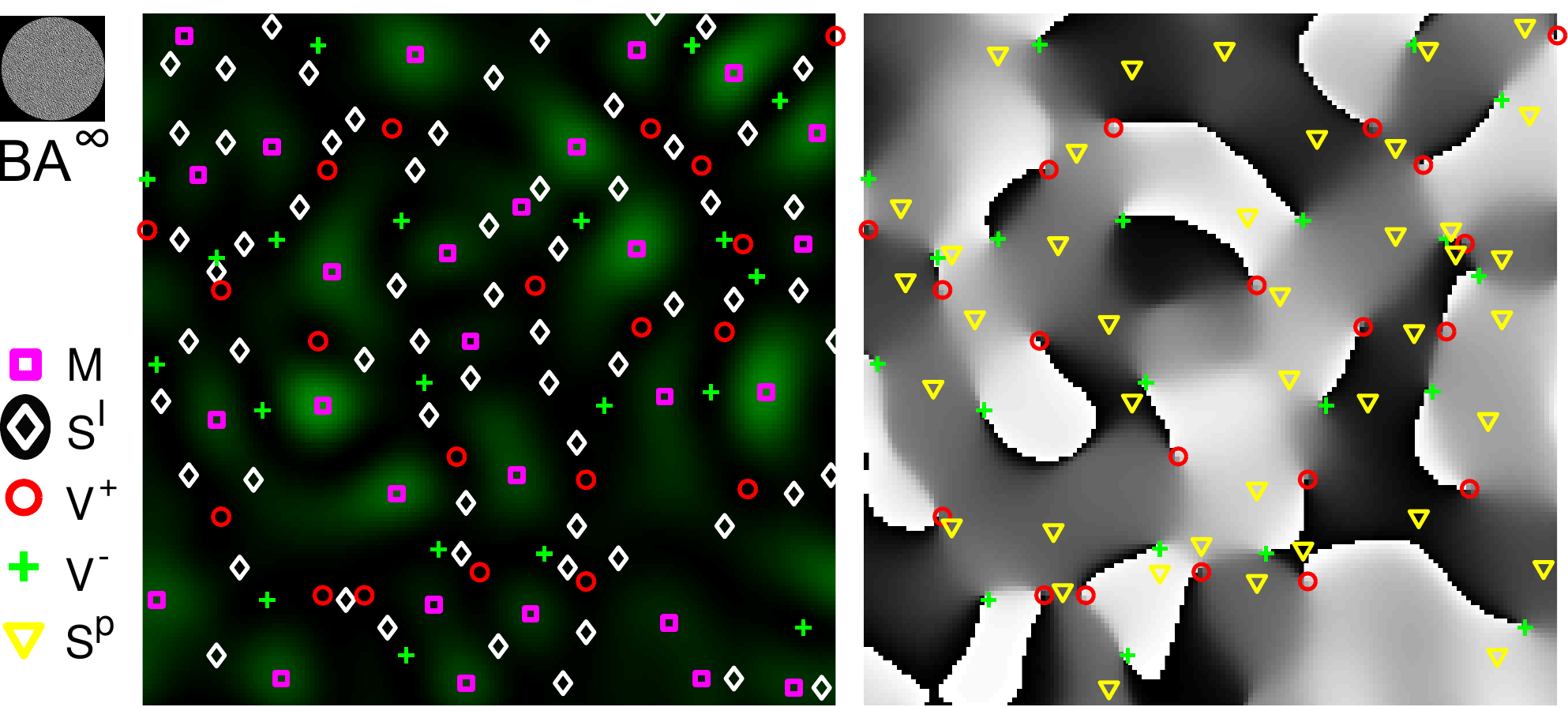}}
\caption{Detected critical points superimposed to the filtered experimental intensity (left) and phase (right) patterns obtained with the aperture $\rm BA^{\infty}$. The displayed pattern correspond to an area of $150\times 150$ pixels on the camera. Phase vortices ($V^+$ and $V^-$) are also zeros of the intensity pattern. Notations for the critical points are identical to those introduced in the core of the main article.}
\label{fig:Expcritpoints}
\end{figure}

The detection of critical points in the experimental speckle pattern is illustrated in Fig.~\ref{fig:Expcritpoints}. The pixel-precise locations of the critical points were determined in the spatial domain using the topology of each pixel neighborhood. For the intensity maxima $M$, the intensity saddles $S^I$ and the phase saddles $S^p$, the eight neighbors of each pixel were used. We used the Matlab function developed by Tristan Ursell and made available in May 2013 (image extrema finder, https://fr.mathworks.com/matlabcentral/fileexchange/41955-find-image-extrema). For the vortices $V^+$ and $V^-$, we used the fact that the summed phase shifts on a closed loop around a vortex is greater than $2\pi$. As a dicretised closed loop, we computed this sum over four neighbouring pixels, the phase differences between the adjacent corners of the square being computed and wrapped in the range $\big] - \pi, \pi \big]$. The summed phase shift was computed by adding the four phase differences, leading to a summed phase shift of $\pm 2\pi$ around a vortex of charge $\pm 1$ and a summed phase shift close to zero elsewhere~\cite{Brauer_91}. 

\section{Numerical simulations of scalar random wavefields} 
\label{sec:Numerical}

Numerical simulations of scalar random wavefields were performed to confirm our experimental results with larger statistics on critical points, independent random phase media and an independent method (section~\ref{sec:vali}). Indeed, larger speckle patterns can be generated numerically directly from randomly distributed phase. Additionally, we extended our investigations to apertures with point group symmetries of orders higher than 4 (section~\ref{sec:highorder}), as well as to apertures with larger angular slits (section~\ref{sec:width}) and to polygon apertures (section~\ref{sec:poly}).

\subsection{Methods for the numerical simulation}

The far-field of uniformly-illuminated apertures comprised of random phases (standard uniform distribution between 0 and $2\pi$,) with spiral phase (${\rm SP}$) masks of order $n \in \llbracket -6;6 \rrbracket$, was computed to simulate the complex wavefield $A_n e^{i\Phi_n}$. The speckle patterns were obtained by computing the two-dimensional Fourier transform of the apertures addressed with random phases. As in the experimental procedure, an uncorrelated speckle pattern was also computed for each aperture. For all the apertures, the speckle grain size (FWHM) was set to $\lambda/(2.NA)= 19$ pixels where $\lambda$ is the optical wavelength and NA is the numerical aperture, and a square grid of 17 mega pixels was computed. The pixel-precise location of the critical points were determined in the spatial domain using the topology of each pixel neighborhood, as for the experimental speckle patterns.

These parameters lead to a count of critical points for the aperture $\rm BA^{\infty}$ of $\sim 10^4$ vortices of each sign. The average number-densities of the critical points for the apertures $\rm BA^{\infty}$, $\rm BA^{3}$ and $\rm BA^{4}$ are presented in Table ~\ref{tab:densitypointssimu}. The slight differences in number-densities between experimental and simulation measurements can be attributed to the higher number of critical points in the simulation, which leads to better precision in the estimation of the critical point densities.

\begin{table}[htbp]
\centering
\caption{\bf Measured average number density of critical points from numerical simulations (length unit: $\lambda/(2.{\rm NA})$). The average number of $V^{-}$ is 9551.43 for the circular aperture (${\rm BA}^{\infty}$). }
\begin{tabular}{ccccc}
\hline
BA mask & $V^{-}$(or $V^{+}$) & $M$ & $S^{p}$ & $S^{I}$  \\
\hline
${\rm BA}^{\infty}$ & $0.20$ & $0.32$ & $0.39$ & $0.67$ \\
${\rm BA}^{3}$ & $0.21$ & $0.38$ &$0.39$ & $0.75$ \\
${\rm BA}^{4}$ & $0.20$ & $0.34$ & $0.43$ & $0.70$ \\
\hline
\end{tabular}
  \label{tab:densitypointssimu}
\end{table}

\subsection{Validation of the experimental results}
\label{sec:vali}

\begin{figure}[htbp]
\centering
\fbox{\includegraphics[width=\linewidth]{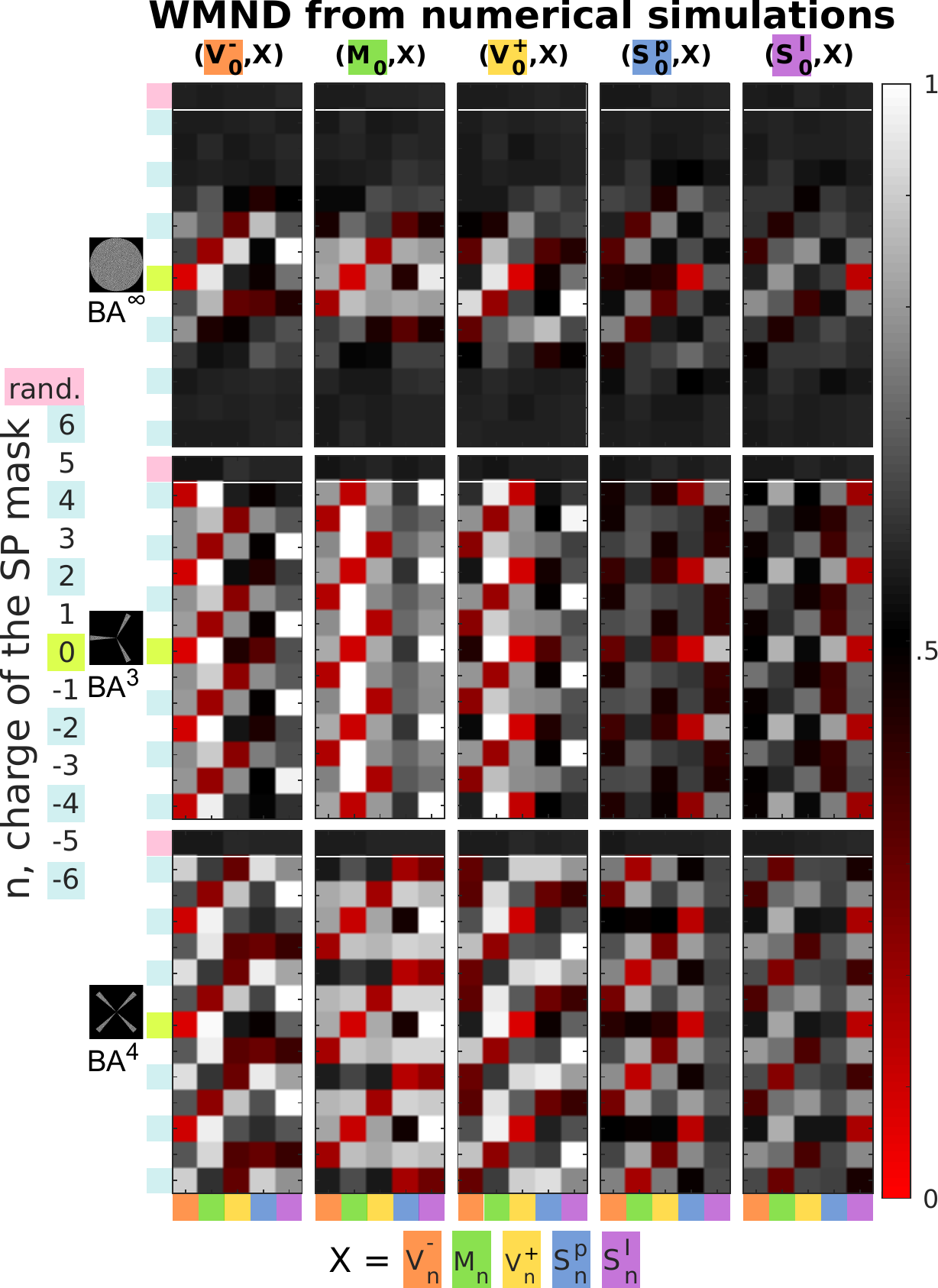}}
\caption{Weighted Median Normalized Distance (WMND) for all possible pairs of critical points screened here, for the addition of spiral phase mask with charges up to n =$\pm 6$ and for the apertures $BA ^{N}$ ($N \in \{ 3,4,\infty \}$. The WMND were computed from numerical simulations.}
\label{fig:simulation34}
\end{figure}

To support our experimental measurements, numerical simulations were performed for the apertures $\rm BA^{\infty}$, $\rm BA^{3}$ and $\rm BA^{4}$. The resulting Weighted Median Normalized Distances (WMND) are presented in Fig.~\ref{fig:simulation34}. Despite the difference in the method to generate the speckle field, the size of the speckle grains, and the number of critical points involved in the statistical estimation of the WMNDs, our experimental results are remarkably consistent with numerical ones. Our experimental results are then validated in two regards. First, the results are robust to modifications in (i) the method to generate the speckle patterns, (ii) the scattering medium and (iii) the spatial sampling. Second, our experimental estimation of the WMND are statistically relevant since a higher average number of critical points did not lead to significantly different WMND.

\subsection{Point group symetries of higher order}
\label{sec:highorder}

\begin{figure}[htbp]
\centering
\fbox{\includegraphics[width=\linewidth]{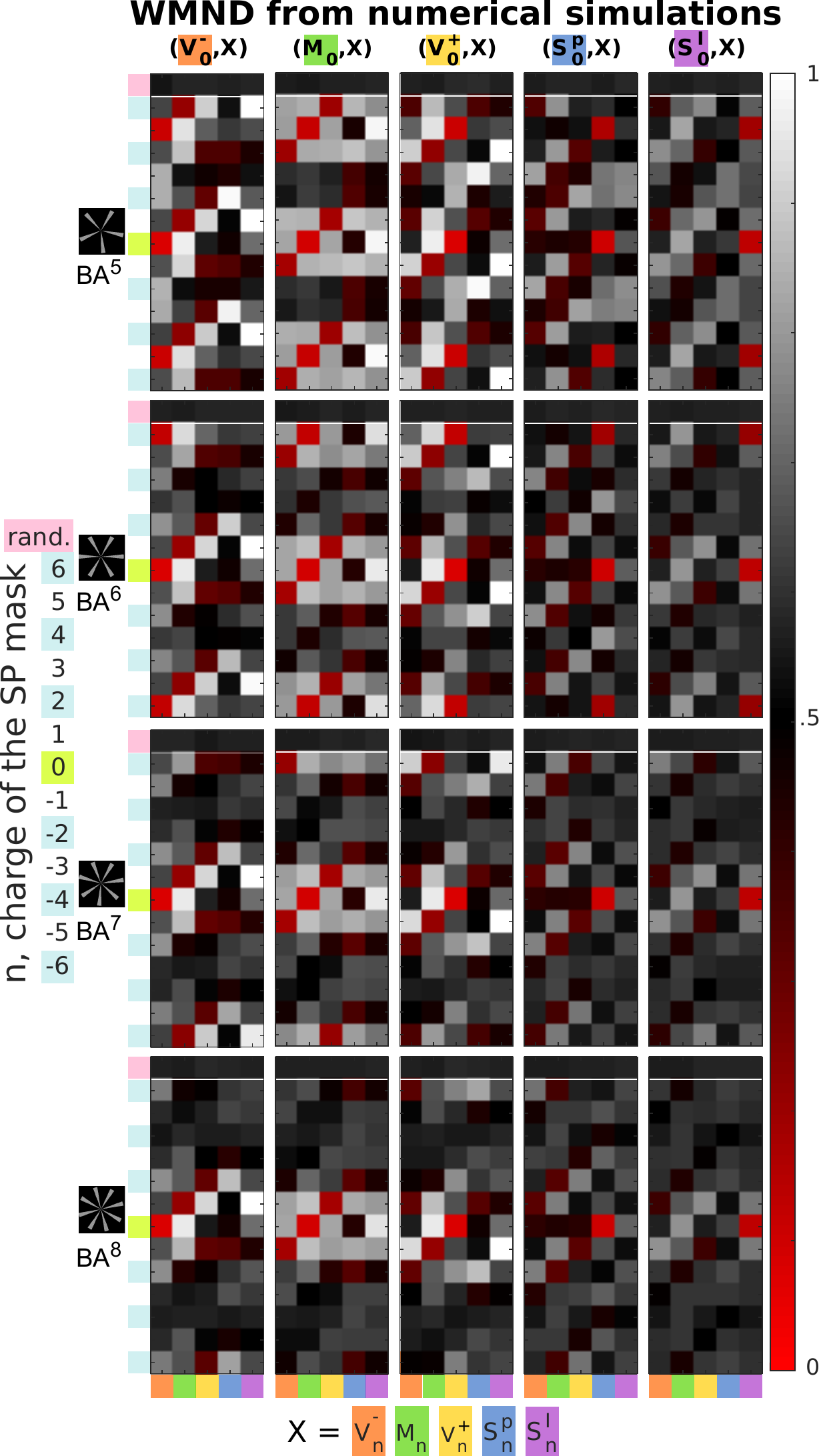}}
\caption{WMND for all possible pairs of critical points screened here, for the addition of spiral phase mask with charges up to n =$\pm 6$ and for the apertures $\rm BA ^{N}$ with $\rm N \in \{ 5,6,7,8 \}$. The WMND were computed from numerical simulations.}
\label{fig:simulation5678}
\end{figure}

We investigated in numerical simulation the effect of an increase in the order of the dihedral point group symmetry. For this purpose, we used the apertures $\rm BA ^{N}$ with $N \in \{ 5,6,7,8 \}$ ( point group symmetry $D_N$), as defined in the main article. Because of the strong anisotropy of the central peak of its point spread function, the aperture $\rm{BA}^2$ could not be compared to the other apertures, and therefore was not computed. The WMND are presented in Fig.~\ref{fig:simulation5678}. 

As expected for N= 5 or 6, the WMND$\left(\rm{Y}_0, \rm{Y}_{\pm N}\right)$ are close to zero (transformation rate close to 1). We can also notice that, for $\rm{BA}^7$ and $n \in \llbracket -3;3 \rrbracket$, as well as for $\rm{BA}^8$ and $n \in \llbracket -4;4 \rrbracket$, the same WMND$\left(\rm{V}_0^+, \rm{X}_{\pm n}\right)$ as for $\rm BA ^{\infty}$ are found. The topological correlations for ${\rm BA}^{\infty}$ correspond to a single spiral mode, while the spiral spectrum of ${\rm BA}^{\rm 8}$ contains harmonics. The first harmonics are the spiral modes $m= \pm 8$ and are for spiral modes high enough so that, for the pairs $\left(\rm{V}_0^+, \rm{X}_{\pm n}\right)$, the topological correlations of the fundamental mode ($m=0$ as in ${\rm BA}^{\infty}$) dominates for $n \in \llbracket -4;4 \rrbracket$. For smaller N, the topological correlations between critical points are dictated by both the fundamental mode and the harmonics which leads to more complex WMND diagrams. This is this true in particular for $\rm BA ^{3}$ and $\rm BA ^{4}$.

\subsection{Influence of the width of the angular slits}
\label{sec:width}

\begin{figure}[htbp]
\centering
\fbox{\includegraphics[width=\linewidth]{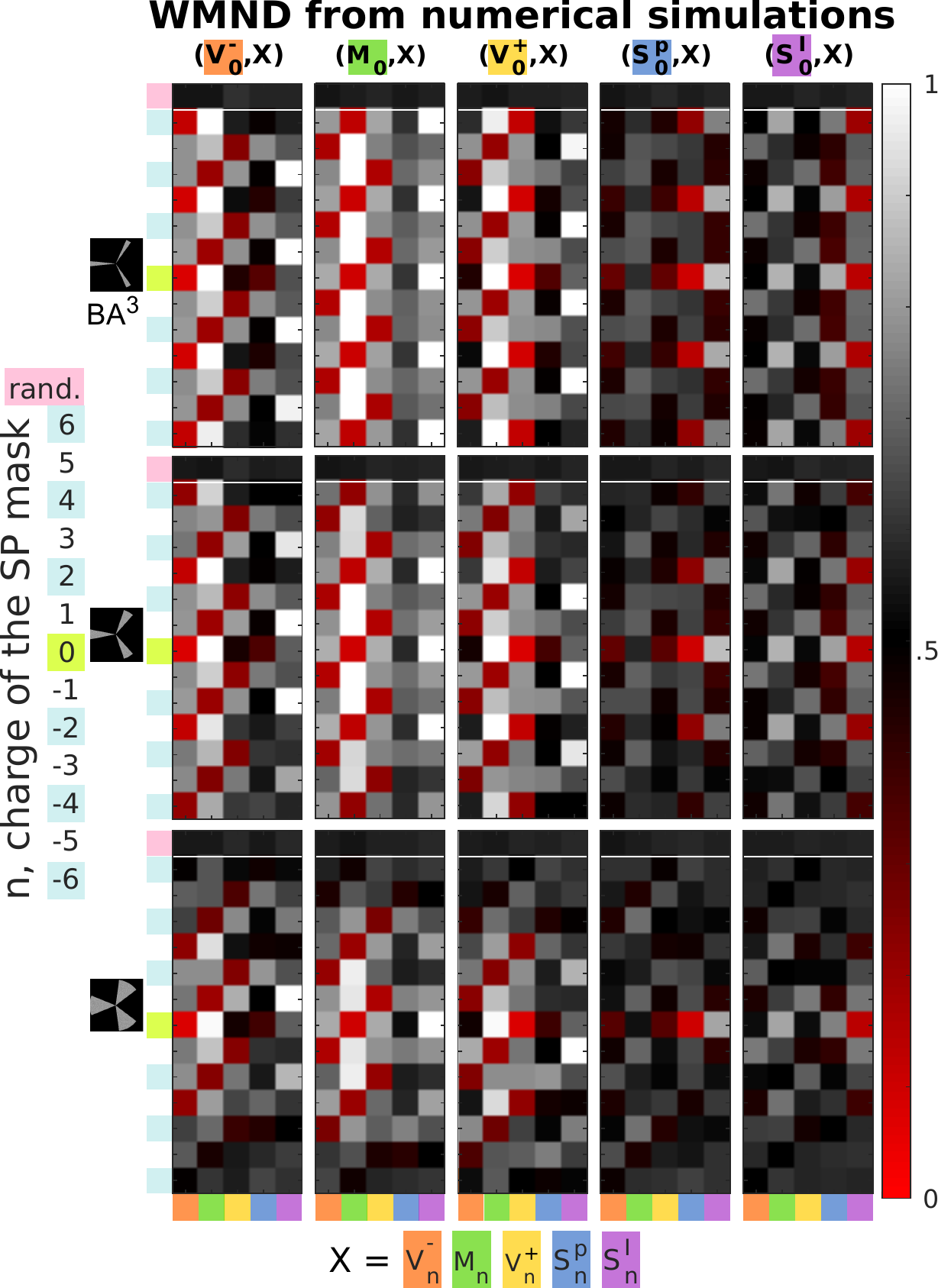}}
\caption{Weighted Median Normalized Distance (WMND) for all possible pairs of critical points screened here, for the addition of spiral phase mask with charges up to n =$\pm 6$ and for apertures with dihedral symmetry $D_3$ and angular slits of width : $\frac{\pi}{16}$, $\frac{\pi}{8}$, and $\frac{\pi}{4}$, respectively (from top to bottom). The WMND were computed from numerical simulations.}
\label{fig:simulationangle}
\end{figure}

The influence of the width of the angular slits was investigated in numerical simulation. Simulation were performed for periodic angular slits with a point group symmetry $D_3$ and three different widths of angular slits were tested : $\frac{\pi}{16}$ (i.e. $\rm BA^3$), $\frac{\pi}{8}$, and $\frac{\pi}{4}$. The WMND are presented in Fig.~\ref{fig:simulationangle}.

We can notice that the WMND$\left(\rm{Y}, \rm{Y}_{\pm 6}\right)$ increases with the increasing width of the slits, and even becomes equal to 0.5 for the largest slits. This results was expected considering the spiral spectrum of the apertures. All the considered apertures have harmonics at the spiral modes $m= k.N$ with $k \in \mathbb{Z}$. However, the amplitude of the harmonics is given by a sinc apodisation function which FWHM depends on the width of the angular slits. For the angular slit $\frac{\pi}{16}$, the FWHM of the spectral apodization function corresponds to n= $ \pm 19$. As a consequence, the aperture could be considered as spectrally invariant by the addition of $\rm SP_{n}$ when $\rm n = \pm 3$ or $\pm 6$. For angular slits of width : $\frac{\pi}{8}$ and $\frac{\pi}{4}$, the FWHM of the spectral apodization function corresponds to n= $ \pm 10$ and n= $ \pm 5$, respectively. Consequently, the apertures are spectrally modified by the addition of $\rm SP_{\pm 6}$. 

In the experimental procedure, the width of the angular slits was chosen large enough so that the dark center induced by the spatial sampling of the SLM pixels has a radius below 2 SLM pixels. However, the width was chosen small enough to keep a spectral invariance in the range $n \in \llbracket -6;6 \rrbracket$. Both the dihedral symmetry and the width of the angular slit are important parmeters to observe the periodicity of the critical point transformation.

\subsection{Polygonal apertures}
\label{sec:poly}

\begin{figure}[htbp]
\centering
\fbox{\includegraphics[width=\linewidth]{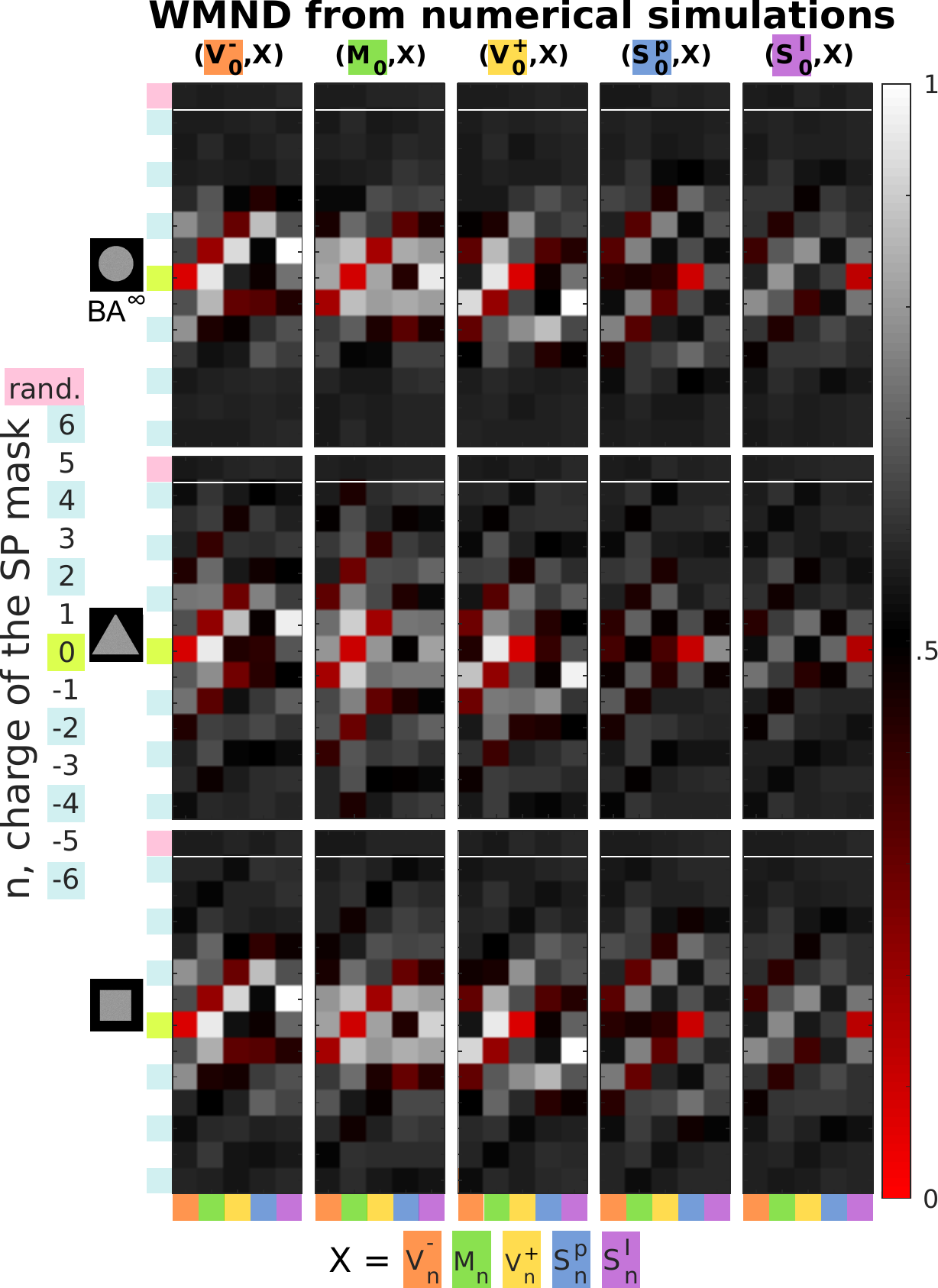}}
\caption{WMND for all possible pairs of critical points screened here, for the addition of spiral phase mask with charges up to n =$\pm 6$ and for a disk aperture (top) and polygonal apertures : equilateral triangle (dihedral symmetry $D_3$) and square (dihedral symmetry $D_4$). The WMND were computed from numerical simulations.}
\label{fig:simulationpoly}
\end{figure}

Polygonal apertures with dihedral symmetries $D_3$ (equilateral triangle) and $D_4$ (square) were used to extend our study to other aperture shapes. The WMND computed from numerical simulations are shown in Fig.~\ref{fig:simulationpoly}.

For the triangular aperture, topological correlations are observed for the pairs $\left(M_0, M_{\pm 3}\right)$, but no cyclic permutation between $V^-$, $M$ and $V+$ can be noticed. For the square, the WMND is very similar to the one of ${\rm BA}^{\infty}$ (disk). 
These results can be explained by the fact that, for N-gons, the $\rm N^{th}$ spiral harmonics have a much lower amplitude than the fundamental spiral mode $(m=0)$. The spiral spectrum is therefore not invariant by the addition of a $\rm SP_{\rm N}$ mask. Moreover, for the triangular aperture, the point spread function (PSF) is an optical lattice which spatial extend expends as the charge of the $\rm SP$ mask increases~\cite{Chavez-Cerda_PRL_10}. Thereby, the maximum values of the PSF decreases, which strongly limits any possible control of the transformation of the critical points.

\end{document}